\documentclass[final,onefignum,onetabnum]{siamart171218}

\usepackage{braket,amsfonts}

\usepackage{array}

\usepackage[caption=false]{subfig}
\usepackage{pgfplots}

\newsiamthm{claim}{Claim}
\newsiamremark{remark}{Remark}
\newsiamremark{hypothesis}{Hypothesis}
\crefname{hypothesis}{Hypothesis}{Hypotheses}

\usepackage{algorithmic}

\usepackage{graphicx,epstopdf}

\Crefname{ALC@unique}{Line}{Lines}

\usepackage{amsopn}

\usepackage{xspace}
\usepackage{bold-extra}
\usepackage{bbold}
\usepackage{float}
\usepackage[most]{tcolorbox}

\colorlet{texcscolor}{blue!50!black}
\colorlet{texemcolor}{red!70!black}
\colorlet{texpreamble}{red!70!black}
\colorlet{codebackground}{black!25!white!25}

\newcommand{\stanx}{\boldsymbol{\mathrm{x}}}

\newcommand{\covariance}[2]{Cov[#1, #2]}

\newcommand{\bb}[1]{\boldsymbol{#1}}

\newcommand{\dashed}[1]{#1^{'}}

\lstdefinestyle{siamlatex}{%
  style=tcblatex,
  texcsstyle=*\color{texcscolor},
  texcsstyle=[2]\color{texemcolor},
  keywordstyle=[2]\color{texemcolor},
  moretexcs={cref,Cref,maketitle,mathcal,text,headers,email,url},
}

\tcbset{%
  colframe=black!75!white!75,
  coltitle=white,
  colback=codebackground, 
  colbacklower=white, 
  fonttitle=\bfseries,
  arc=0pt,outer arc=0pt,
  top=1pt,bottom=1pt,left=1mm,right=1mm,middle=1mm,boxsep=1mm,
  leftrule=0.3mm,rightrule=0.3mm,toprule=0.3mm,bottomrule=0.3mm,
  listing options={style=siamlatex}
}

\newtcblisting[use counter=example]{example}[2][]{%
  title={Example~\thetcbcounter: #2},#1}

\newtcbinputlisting[use counter=example]{\examplefile}[3][]{%
  title={Example~\thetcbcounter: #2},listing file={#3},#1}

\DeclareTotalTCBox{\code}{ v O{} }
{ 
  fontupper=\ttfamily\color{black},
  nobeforeafter,
  tcbox raise base,
  colback=codebackground,colframe=white,
  top=0pt,bottom=0pt,left=0mm,right=0mm,
  leftrule=0pt,rightrule=0pt,toprule=0mm,bottomrule=0mm,
  boxsep=0.5mm,
  #2}{#1}

\patchcmd\newpage{\vfil}{}{}{}
\begin{tcbverbatimwrite}{tmp_\jobname_header.tex}
\title{Diagnostics-Driven Nonstationary Emulators Using Kernel Mixtures\thanks{Submitted to the editors DATE.
\funding{Daniel Williamson was funded by EPSRC fellowship (EP/K019112/1) and by the NERC EuroClim project (NE/M006123/1). }}}

\author{Victoria Volodina
\and Daniel B. Williamson\thanks{College of Engineering, Mathematics and Computer Sciences, University of Exeter}}

\headers{Diagnostics-Driven Nonstationary Emulators Using Kernel Mixtures}{Victoria Volodina and Daniel B. Williamson}
\end{tcbverbatimwrite}
\input{tmp_\jobname_header.tex}

\ifpdf
\hypersetup{ pdftitle={Guide to Using  SIAM'S \LaTeX\ Style} }
\fi

\begin{document}
\maketitle

\begin{tcbverbatimwrite}{tmp_\jobname_abstract.tex}
\begin{abstract}
Weakly stationary Gaussian processes (GPs) are the principal tool in the statistical approaches to the design and analysis of computer experiments (or Uncertainty Quantification). Such processes are fitted to computer model output using a set of training runs to learn the parameters of the process covariance kernel. The stationarity assumption is often adequate, yet can lead to poor predictive performance when the model response exhibits nonstationarity, for example, if its smoothness varies across the input space. In this paper, we introduce a diagnostic-led approach to fitting nonstationary GP emulators by specifying finite mixtures of region-specific covariance kernels. Our method first fits a stationary GP and, if traditional diagnostics exhibit nonstationarity, those diagnostics are used to fit appropriate mixing functions for a covariance kernel mixture designed to capture the nonstationarity, ensuring an emulator that is continuous in parameter space and readily interpretable. We compare our approach to the principal nonstationary GP models in the literature and illustrate its performance on a number of idealised test cases and in an application to modelling the cloud parameterization of the French climate model.
\end{abstract}

\begin{keywords}
Emulation, Nonstationary model response, Bayesian Methods, Uncertainty Quantification, Diagnostics
\end{keywords}

\begin{AMS}
  62P12, 97K80, 60G15, 62F15, 62J20
\end{AMS}
\end{tcbverbatimwrite}
\input{tmp_\jobname_abstract.tex}

\section{Introduction}
\label{sec:intro}
Computer models are widely used to model physical processes \cite{Frierson2006,Farneti2009}. These models are treated as mathematical functions, denoted $f$, of a large number of parameters, denoted $\stanx$, and are often computationally expensive to run. Gaussian Process (GP) emulators are often used as `surrogates' to complex computer models in such cases, in order to make inference that requires embedding the model as part of a Monte Carlo procedure, tractable. The usual approach to emulation is to fit a stationary GP \cite{Sacks1989,Kennedy2001,Santner2003} trained on a set of runs of the computer model. GP emulators are then used in a number of inferential engines, such as uncertainty analysis \cite{Oakley2002}, sensitivity analysis \cite{Oakley2004} and calibration \cite{Kennedy2001}.

The usual stationarity assumption gives the process a covariance function of the form \[\covariance{f(\stanx)}{f(\stanx')} = \sigma^2r(|\stanx-\stanx'|,\bb{\delta})
\]
where $\sigma^2$ is a common variance term and $r(\cdot)$ is a correlation function that depends only on the distance between $\stanx$ and $\stanx'$ and parameters $\bb{\delta}$ \cite{Bastos2009}. However, in practice, computer models may have different covariance structures across the input space. This can sometimes be captured by using complex regression functions in the prior mean of the GP so that the residual is approximately stationary \cite{Rougier2009, Williamson2013}. When the nonstationarity is not adequately captured in this way, prediction intervals produced by the emulators can be too narrow in the region of high residual variability of $f$ (the emulator is over-confident). On the contrary, the emulator is under-confident in the input space where $f$ is `well-behaved' \cite{Bastos2009, Montagna2016, Oughton2016}. 

There is a large literature on nonstationary GP models in spatial statistics. For example, \cite{Gelfand2003} suggested modelling spatial processes through the mean function using spatially varying regression coefficients, with the coefficients themselves given a multivariate process model. \cite{Paciorek2006} proposed to use nonstationary GPs for spatial modelling, by convolving spatially-varying kernel functions. These methods have not yet been implemented for computer experiments as it would take many more model runs than are usually possible to obtain designs that are sufficiently dense in the typically $> 2$ input dimensions that we work with. 

A number of approaches that aim to model a nonstationary response with modified GP models have appeared in the Uncertainty Quantification (UQ) literature. One idea is to find a nonlinear transformation of the input space to achieve stationarity \cite{Sampson1992,Schmidt2003,Ginsbourger2017}. Several approaches aim to reduce the computational burden for large computer models and provide nonstationary modeling features at the same time. Examples include sparse pseudo-inputs \cite{Snelson2006} and local GPs \cite{Gramacy2015}. Recently, \cite{Montagna2016} presented a nonstationary GP by augmenting the parameter space with a latent GP. The latent input is used to capture regions in the input space that correspond to abrupt changes in the simulator behaviour. Collectively, we might view these approaches as generally warping the input space in some way to achieve nonstationarity.

Another general approach is to partition the input space and to fit separate GPs for each partition. \cite{Gramacy2008} used treed partitioning and make splits on the value of a single variable. An extension to treed partitioning is Voronoi tessellation that uses the Euclidean distance from a set of centers to create Voronoi cells \cite{Kim2005, Pope2018}.

Several works aim to deal with the nonstationarity through separating the model response in terms of a global, large-scale behaviour, and a locally stationary process. One example is to fit a complex mean function using relatively dense polynomials and then to capture the residual via a stationary GP \cite{Rougier2009,Vernon2010,Williamson2013}. The composite Gaussian Process (CGP) \cite{Ba2012} replaces the polynomial mean function with another GP with a longer length scale to capture the global trend throughout the input space. A flexible variance model is introduced to capture varying volatility across the input space.

We are interested in developing nonstationary GPs which retain interpretability so that prior elicitation for their parameters is possible, when sufficient expert knowledge exists regarding the function behaviours, or so that the right type of nonstationary GP can be fitted in response to a standard stationary GP failing the usual diagnostic checks. Whilst the warping type methods we referred to above can be powerful for certain model output, the structure and posteriors over the parameters can be difficult to interpret. We do view approaches to fit complex ``global" mean functions to be appropriate as they offer interpretability, but still often find more complex nonstationarity in the residual so that methods like CGP are not as effective. Partitioning approaches can be very effective as, when fitting diagnostics, we often ``see" regions of different behaviour, and it can be intuitive for the modellers to think of the model having different characteristics in different input regions. However, we do not believe the implied boundary discontinuities that these models specify. 

The method we propose here develops a kernel mixture approach to the GP residual offering a flexible and interpretable model. Our approach has similarities to an approach from the spatial statistics literature. \cite{Fuentes2001} and \cite{Banerjee2004} assume the existence of $L$ stationary processes in different regions of a 2D spatial field and specify $L$ centroids in that field by applying rectangle-partitioning. Together with a weight function dictated by the distance between the point of interest and the centroids, they specify a nonstationary kernel for the whole space as the weighted sum of $L$ stationary region-specific covariance kernels. 

In this paper, and in the absence of model-specific expert knowledge regarding potential model input space regions and different properties within them, we partition the input space using diagnostics from initial stationary GP fits to develop a single nonstationary GP emulator with a flexible mixture kernel obtained via the partition. Firstly, we perform standardized Leave One Out (LOO) diagnostics for a stationary GP emulator fitted to our training data. We specify a finite mixture model for the standardized LOO errors to identify $L$ distinct regions of behaviour in the input space. We then assign a Gaussian process for $f$ whose covariance kernel is a mixture of $L$ stationary covariance kernels, each belonging to the $L$ previously identified regions. This allows us to fit a single Gaussian process and operate within the original input space.

The article has the following structure. \Cref{sec:GPE} gives an overview of the stationary GP emulator and describes the existing work in nonstationary GP modelling in more detail. \Cref{sec:NGPE} introduces our nonstationary GP emulator.   
\Cref{sec:case} examines the performance of our nonstationary GP emulator for a number of idealised numerical examples. In \cref{sec:cnrm} we apply our methodology to the boundary layer of the single column ARPEGE-Climat model. In particular, the climate modellers we collaborate with are interested in calibrating cloud parameterizations within their global models so that the sub-grid scale processes they represent have similar performance to models that explicitly resolve these processes. During our work with this project we observed that some of the model outputs we needed to capture failed traditional diagnostics checks, motivating this work. \Cref{sec:disc} contains a discussion.

\section{Gaussian process emulators}
\label{sec:GPE}
\subsection{Stationary GPs}
\label{subsec:STGP}
We define a computer model to be a function, $f$, that takes inputs $\stanx \in \mathcal{X}$ and outputs a scalar. Multivariate emulation often uses the scalar output emulator as a building block \cite{Higdon2008,Bayarri2009,Salter2018,Rougier2008}.
We write the GP emulator as
\begin{equation} \label{eq:1}
f(\stanx) \vert \bb{\beta}, \sigma^2, \bb{\delta}, \tau^2 \sim GP \big(h(\stanx)^T \bb{\beta}, k(\cdot , \cdot ; \sigma^2, \bb{\delta}, \tau^2) \big), 
\end{equation}
with regression functions $h(\stanx)$, coefficients $\bb{\beta}$, and covariance
\[ Cov[f(\stanx), f(\stanx')] = k(\stanx, \stanx'; \sigma^2, \bb{\delta}, \tau^2) = \sigma^2r(\vert\stanx-\stanx'\vert,\bb{\delta}) + \tau^2 \mathbb{1}\big\{ \stanx = \stanx'\big\}, \]
where $\sigma^2$ and $\tau^2$ are variance parameters and $\bb{\delta}$ are parameters of the correlation function $r(\cdot,\cdot)$.
The indicator function $\mathbb{1}\big\{ \stanx = \stanx'\big\}$ is defined as 
\begin{equation*}
\mathbb{1}\big\{\stanx=\stanx' \big\} = \begin{cases} 1, & \stanx = \stanx',\\
0, & \stanx \neq \stanx'.
\end{cases}
\end{equation*}
There are a number of popular forms of correlation function leading to positive definite kernel, $k$, used routinely in the literature on computer experiments, such as the power exponential
\[ r(\vert\stanx-\stanx'\vert, \bb{\delta})= \exp \Big\{ -\sum_{j=1}^{p} \Big(\frac{x_j - x'_j}{\delta_j}\Big)^{\phi_j}\Big\} , \]
with $0< \phi_j\leq 2$, and $\phi_j = 2$ corresponding to the popular squared exponential. The Matern covariance function is also commonly used \cite{Santner2003, Rasmussen2004}. \par 
 We can therefore consider $f(\stanx)$ to be the sum of 3 processes
\begin{equation}
f(\stanx) = h(\stanx)^T\bb{\beta} + \epsilon(\stanx) + \nu(\stanx)
\end{equation}
where $h(\stanx)^T\bb{\beta}$ represents a global response surface that captures dominant features of the model output, $\epsilon(\stanx)$ is a correlated residual process capturing local input dependent deviation from the global response surface and $\nu(\stanx)$ is a nugget process either representing noise in the simulator (such as the internal variability of a climate model \cite{Williamson2017}) or for capturing variability due to inactive inputs not used in $\epsilon(\stanx)$ \cite{Andrianakis2012}, or to ensure non-singularity of covariance matrices in the Bayesian update \cite{Gramacy2012}. 

Define $\bb{X} = (\stanx_1, \dots, \stanx_n)^T$ at which we obtain training runs $\bb{F} = (f(\stanx_1), \dots, f(\stanx_n))$. Conditioned on the hyperparameters, design $\bb{X}$ and $\bb{F}$, we perform the Bayesian update to derive the posterior distribution for $f(\stanx)$.
\begin{equation}
f(\stanx) \vert \bb{X}, \bb{F}, \bb{\beta}, \sigma^2, \bb{\delta}, \tau^2 \sim GP\big(m^{*}(\stanx), k^{*}(\cdot, \cdot)\big)
\end{equation}
with 
\[ m^{*}(\stanx) = h(\stanx)^T\bb{\beta}+k(\stanx, \bb{X})\bb{K}^{-1} (\bb{F}-h(\stanx)^T\bb{\beta})\]
\[ k^{*}(\stanx, \stanx') = k(\stanx, \stanx') - k(\stanx, \bb{X})\bb{K}^{-1} k(\bb{X}, \stanx')\]
and $\bb{K}=k(\bb{X}, \bb{X})$, with $\bb{K}_{ij} = k(\stanx_i, \stanx_j)$. 

There are a number of popular methods for fitting GPs. \cite{Currin1991} fit the hyperparameters by maximum likelihood. \cite{ Haylock1996} showed that fixing $\bb{\delta}$ first led to a t-process posterior prediction for $f(\stanx)$ under $\pi(\bb{\beta}, \sigma^2) \propto 1/\sigma^2$. This was extended to a Normal-inverse gamma update by \cite{Oakley2004}, given $\bb{\delta}$. Full Bayes MCMC methods with specific priors on $\bb{\delta}$ have been used \cite{Higdon2008, Kaufman2010} and will be used in this paper. 

\begin{figure}[ht]
\begin{center}
\includegraphics[height=0.2\textheight, width=0.6\textwidth]{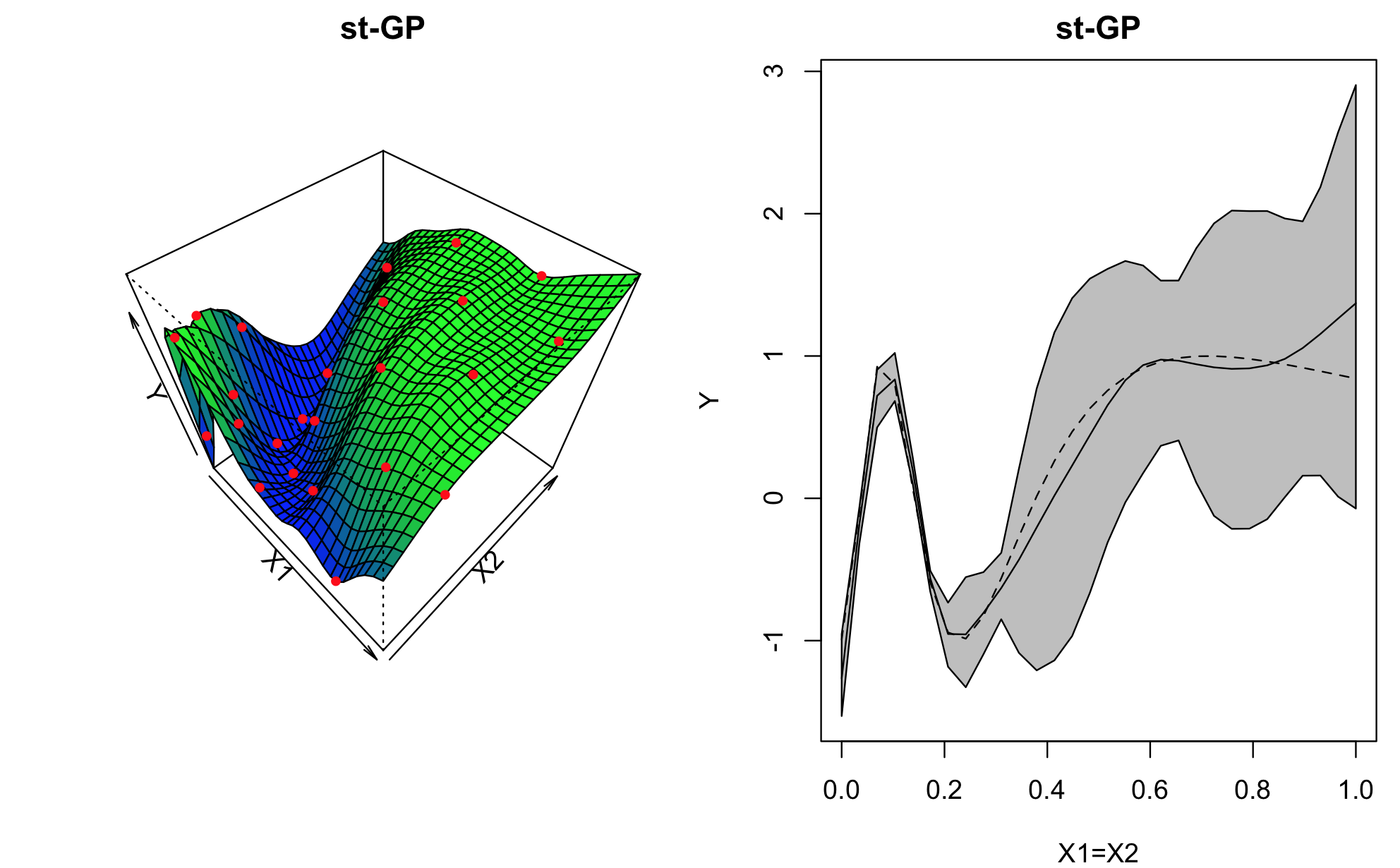}
\includegraphics[height=0.2\textheight, width=0.6\textwidth]{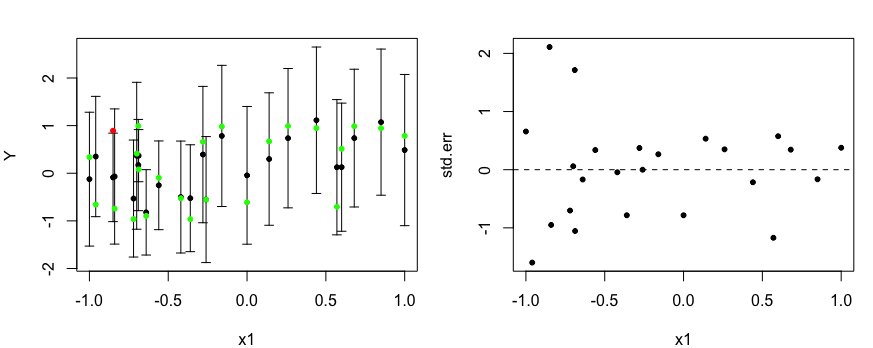}
\end{center}
\caption{Failure of stationary GP emulator. \textit{Top left}: Posterior mean predictive surface produced by stationary GP emulator with 24 design points in red. \textit{Top right}: Emulator performance for the cross section $\mathrm{x}_1=\mathrm{x}_2$. The dashed line is the true function values, the solid black line is the posterior mean predictive curve, and the grey areas denote two standard deviation prediction intervals. \textit{Bottom left}: Leave One Out diagnostic plot against $\mathrm{x}_1$. The predictions and two standard deviation prediction intervals are in black. The true function values are in green if they lie within two standard deviation prediction intervals, or red otherwise. \textit{Bottom right}: Individual standardized errors of emulator predictions against $\mathrm{x}_1$.}
\label{fig:figure1}
\end{figure}

In practice even with very flexible or well-chosen prior distributions, there can be difficulties with stationary emulators for highly nonstationary simulators.  \cref{fig:figure1} demonstrates the performance of a stationary GP emulator for a 2D function considered by Ba and Joseph in the \textsf{R} package \textsf{CGP}'s reference manual \cite{Ba2018}. From the top left panel, the mean surface is too rough in the highly smooth area for $\mathrm{x}_1$ and $\mathrm{x}_2$ large. We also see overconfidence in the region where $f$ fluctuates rapidly, for small values of $\mathrm{x}_1$ and $\mathrm{x}_2$,  and underconfidence in the region where the function is relatively smooth, for large values of $\mathrm{x}_1$ and $\mathrm{x}_2$ from the top right panel. The diagnostic plots recommended in \cite{Bastos2009} shown in the lower left and right panels of \cref{fig:figure1} are indicative of problems we commonly encounter when building emulators in practice. The leave one out cross validation plot (bottom left) shows that a calibration using this emulator would be inefficient in identifying optimal regions of parameter space \cite{Kennedy2001, Salter2018}, and the heteroskedasticity in the standardized errors plot (bottom right) is a typical issue. 

\subsection{Existing work in nonstationary GP modelling}
\label{subsec:LitReview}
There are a number of approaches that adapt GPs to model nonstationary functions. Spatial deformation approaches include \cite{Sampson1992}, \cite{Schmidt2003} and \cite{Ginsbourger2017}. \cite{Sampson1992} use a multidimensional scaling (MDS) representation of the original inputs to achieve nonstationarity. The smooth mapping to MDS from original input space is achieved through thin-plate splines. \cite{Schmidt2003} use a Gaussian Process prior for a mapping from the original input space to the transformed stationary input space. \cite{Ginsbourger2017} combines dimensional reduction with a multiple index model and a non-linear mapping in the input space to achieve a nonstationary GP.

Several approaches aim to model large computer experiments and provide the flexibility to model nonstationarity in the response as well. 
For example, \cite{Snelson2006} introduce Sparse Pseudo-input Gaussian Processes (SPGPs) by defining pseudo-inputs (pseudo-design), $\bar{\bb{X}}=(\bar{\stanx}_1, \dots, \bar{\stanx}_M)$ with $M<n$, and the pseudo targets, $\bar{\bb{f}}=(\bar{f}_1, \dots, \bar{f}_M)$, which are zero mean Gaussian with covariance $\bb{K}_M=k(\bar{\bb{X}}, \bar{\bb{X}})$. The Gaussian Process predictive distribution is parameterized by a pseudo data set and we derive the following form of the complete data likelihood
\[p(\bb{F}\vert \bb{X}, \bar{\bb{X}}, \bar{\bb{f}})=N\big(\bb{K}_{NM}\bb{K}_M^{-1}\bar{\bb{f}}, \bb{\Lambda}+\sigma^2\bb{I} \big), \]
where $\bb{K}_{NM}=k(\bb{X}, \bar{\bb{X}})$ and $\bb{\Lambda}=diag(\bb{\lambda}),$ with $\lambda_i=k(\stanx_i, \stanx_i)-k(\stanx_i, \bar{\bb{X})}\bb{K}_M^{-1}k(\bar{\bb{X}}, \stanx_i)$ for $i=1, \dots, n$.
The pseudo-inputs are treated the same way as GP model hyperparameters and are derived by maximizing marginal likelihood $p(\bb{F}\vert \bb{X}, \bar{\bb{X}}, \bb{\beta}, \sigma^2, \bb{\delta})$ with respect to $\big\{\bar{\bb{X}}, \bb{\beta}, \sigma^2, \bb{\delta}\big\}$ by gradient descent. SPGP allows to model nonstationarity by moving the pseudo-inputs to the interesting parts of input space. Another approach, Local Gaussian Processes \cite{Gramacy2015} focuses on the prediction problem for $\stanx$ locally. Instead of dealing with the full design set $\bb{X}$, the design is derived sequentially by considering the mean-squared predictive error at $\stanx$. Operating with design locally, relative to $\stanx$, allows users to deal with nonstationarity in the model response by focusing on local model behaviour. 

A common approach in mitigating nonstationarity is to fit a more complicated response surface, $h(\stanx)^T\bb{\beta}$, that captures global nonstationarity, leaving a stationary process residual \cite{Rougier2009, Vernon2010, Williamson2013}. If the right functions can be found and added to $h(\stanx)$, they should be used, but in practice this can be extremely difficult and can require too much manual fitting. In applications where emulators are needed quickly or automatically (say 100's or 1000's must be built to describe a complex computer code), this may not be feasible in such a way as to leave a stationary residual. 

\cite{Ba2012} provides an alternative to fitting a complex mean function with the Composite Gaussian Process (CGP). CGP consists of two processes, 
\begin{equation}
f(\stanx) = Z_{global}(\stanx) + \sigma(\stanx)Z_{local}(\stanx).
\end{equation}

They specify $Z_{global}(\stanx)$ is GP distributed with mean $\mu$ and covariance $\tau^2g(\stanx, \stanx')$ and $Z_{local}(\stanx)$ is zero-mean GP with covariance $l(\stanx, \stanx').$ The first process is used to model the global trend, while the second process captures the local variability by including a variance model $\sigma^2(\stanx)=\sigma^2 v(\stanx)$, where $\sigma^2$ is a variance term and $v(\stanx)$ is a standardized volatility function which fluctuates around 1. Both $g(\stanx,  \stanx')$ and $l(\stanx, \stanx')$ are the Gaussian correlation functions:
\begin{equation}
g(\stanx, \stanx')=\exp\Big(-\sum_{k=1}^p\theta_k(\mathrm{x}_{k}-\mathrm{x}'_{k})^2\Big), \quad l(\stanx, \stanx')=\exp\Big(-\sum_{k=1}^p \alpha_k(\mathrm{x}_{k}-\mathrm{x}'_{k})^2 \Big) \nonumber
\end{equation}
with their own correlation parameters that satisfy $\bb{0}\leq \bb{\theta} \leq \bb{\alpha^l}$ and $\bb{\alpha^l}\leq \bb{\alpha}$. The final model specification is $f(\stanx)\sim GP\Big(\mu, \tau^2g(\cdot) + \sigma^2v(\stanx)l(\cdot) \Big)$, which could be considered as a GP model with flexible covariance structure. 

Alternatively, several works in literature use piecewise GPs to model nonstationarity. \cite{Gramacy2008} developed a treed partition model to divide the input space by making binary splits on the value of single variables recursively. The Bayesian approach is applied to the partitioning process by specifying a prior through a tree-generating process and the tree is averaged out by integrating over possible trees using Reversible Jump MCMC (RJ-MCMC). In each leaf of the tree a stationary Gaussian Process is fitted. Partitioning does not guarantee continuity of the fitted function, because the posterior predictive surface conditional on a particular tree is discontinuous across the partition boundaries. This translates into higher posterior predictive uncertainty near region boundaries. However, Bayesian model averaging provides mean fitted functions that are relatively smooth in practice. \cite{Pope2018} provided an extension to these classification trees for modelling discontinuities in the model response. Voronoi tessellation is applied to divide the input space into disjoint regions and an independent Gaussian Process is fitted to each region, with RJ-MCMC used for implementation.
\section{Nonstationarity via diagnostic-driven covariance mixtures}
\label{sec:NGPE}
When fitting an emulator in practice, we would normally begin by fitting a stationary Gaussian Process and examining diagnostics to assess whether the emulator was sufficient. Possible failure of the stationarity assumption can then be checked from the plots of standardized errors against the model inputs \cite{Bastos2009}. We may notice, as we do in \cref{fig:figure1}, that the model is `well-behaved' in some regions of the input space but not in others. For example, the standardized errors are close to zero for $\mathrm{x}_1$ and $\mathrm{x}_2$ close to 1 in \cref{fig:figure1}, yet the model changes rapidly and the standardized errors are large for small values of $\mathrm{x}_1$ and $\mathrm{x}_2$. Approaches such as TGP explicitly model these as regions of distinct behaviour by axis-aligned partitioning of the input space and fitting distinct Gaussian Processes to each region. Our approach captures the distinct regional behaviours we see in stationary diagnostics, yet uses input-dependent mixing functions to ensure a continuous covariance kernel. We develop this approach below. 
\subsection{Nonstationary GP through mixtures of stationary processes}
\label{sec:ProposeMethod}
Suppose, upon examining the diagnostics of a stationary GP emulator, as above, we identify $L$ input regions of distinct model behaviour, $\mathcal{X}_l$, $l=1, \dots, L$ (see \cref{sec:Mixture} for our method for identifying these regions and the optimal number of these regions). We define $f(\stanx)$ as:
\begin{equation}
\label{eq: eqmod}
f(\stanx) = h(\stanx)^T\bb{\beta} + \sum_{l=1}^L \lambda_l(\stanx) \epsilon_l(\stanx) +\sum_{l=1}^L z_l(\stanx)\nu_l(\stanx),
\end{equation}
where $\lambda_1(\stanx), \dots, \lambda_L(\stanx)$ are input-dependent mixture components on the unit simplex, i.e. $\sum_{l=1}^L \lambda_l(\stanx)=1$. Here $\epsilon_l(\stanx)$ are independent, mean zero, Gaussian processes with covariance kernel $k_l(\cdot,\cdot; \sigma^2_l, \bb{\delta}_l)$, so that
\[ \epsilon_l(\stanx) \vert \sigma_l^2, \bb{\delta}_l \sim GP\Big(0, k_l(\cdot, \cdot; \sigma_l^2, \bb{\delta}_l)\Big).\]
The final term of equation \ref{eq: eqmod} is a nugget process term. We specify a nugget process term for each input region, allowing the nugget to vary between the regions
\[\nu_l(\stanx) \sim N(0, \tau_l^2),\quad z_l(\stanx) = \mathbb{1}\big(\lambda_l(\stanx) = \max_k \big\{ \lambda_k(\stanx)\big\} \big). \]
We define $z_l(\stanx)$ as the indicator function of the form
\begin{equation*}
z_l(\stanx) = \begin{cases}
1, &\lambda_l(\stanx) = \max_{k}\big\{\lambda_k(\stanx) \big\},\\ 
0, &\mbox{otherwise},
\end{cases}
\end{equation*}
which allows us to use a single nugget process term from one of the regions $l=1, \ldots, L$, in equation \ref{eq: eqmod} by finding $l$ that provides us with the maximum from the mixture components, $\lambda_1(\stanx), \ldots, \lambda_L(\stanx)$, evaluated at the point of interest $\stanx$. This specification allows nugget process to be region specific but it does not vary in the same way as the residual process \cite{Banerjee2004}. We mentioned in \cref{subsec:STGP} that nugget term could account for the noise in the simulator output or the effect of inactive inputs in the residual term and we consider it as unstructured term in our model. 

Given $\bb{\lambda}(\stanx) = (\lambda_1(\stanx), \cdots, \lambda_L(\stanx))$ and region-specific parameters $\Delta=(\bb{\delta}_1, \dots, \bb{\delta}_L)^T$, $\bb{\sigma}^2=(\sigma_1^2, \dots, \sigma_L^2)$ and $\bb{\tau}^2 = (\tau_1^2, \dots, \tau_L^2)$ our nonstationary GP is therefore
\begin{equation}
f(\stanx) \vert \bb{\beta}, \bb{\lambda}(\stanx), \bb{\sigma}^2, \Delta, \bb{\tau}^2 \sim GP \Big(h(\stanx)^T\bb{\beta}, k(\cdot, \cdot; \bb{\sigma}^2, \Delta, \bb{\tau}^2) \Big)
\end{equation}
with 
\begin{equation}
\label{eq:covarnonstat}
k(\stanx, \stanx'; \bb{\sigma}^2, \Delta, \bb{\tau}^2) = \sum_{l=1}^L\lambda_l(\stanx)\lambda_l(\stanx')k_l(\stanx, \stanx'; \sigma_l^2, \bb{\delta}_l) + \mathbb{1}\big\{ \stanx = \stanx'\big\} \sum_{l=1}^L z_l(\stanx) z_l(\stanx')\tau_l^2, 
\end{equation}
so that the covariance kernel for our nonstationary GP is a mixture of stationary covariance kernels. This formulation allows us to specify a different type of process behaviour in $L$ regions similar to TGP. However by mixing GPs in this way we can have non-zero covariance between the points from different regions. 
\subsection{Modelling the mixture components}
\label{sec:Mixture}
We consider a vector of probabilities, $\bb{\lambda}(\stanx)=(\lambda_1(\stanx), \dots, \lambda_L(\stanx))$, indicating the dominant local behaviour around $\stanx$ in each region $l=1, \dots, L$ as described by $\epsilon_l(\stanx)$ and $\nu_l(\stanx)$. Define $\tilde{\bb{f}}(\stanx)$ to be an `out of the box' stationary emulator fitted using design $\bb{X}$ and computer model runs $\bb{F}$. Let $e_i$ denote the leave-one-out standardized cross validation residuals: 
\begin{equation}
e_i = \frac{f(\stanx_i)-E[\tilde{\bb{f}}(\stanx_i)\vert\bb{X}_{-i}, \bb{F}_{-i}]}{SD[\tilde{\bb{f}}(\stanx_i)\vert \bb{X}_{-i}, \bb{F}_{-i}]} \nonumber
\end{equation}
where $E[\cdot]$ and $SD[\cdot]$ denote the expectation and standard deviation respectively evaluated at $\stanx$. Closed form formulas are available for fast computation of $E[\tilde{f}(\stanx_i)\vert \bb{X}_{-i}, \bb{F}_{-i}]$ and $SD[\tilde{f}(\stanx_i)\vert \bb{X}_{-i}, \bb{F}_{-i}]$ \cite{Bachoc2013}.

We define a latent indicator process $s(\stanx)$ as
\begin{equation}
 s(\stanx) \sim \text{Multinomial}(\lambda_1(\stanx), \dots, \lambda_L(\stanx) ),  \nonumber
 \end{equation}
 and model the $e_i$ via
 \[ e_i \vert s(\stanx_i) =l \sim \text{Normal}(0, \zeta_l), \]
 where $\zeta_l$ is the standard deviation for the distribution of standardized errors in region $l$. This particular specification comes from the model diagnostic, i.e. we adopt a strong assumption that the standardized errors are standard normally distributed \cite{Bastos2009}. The patterns of small and large errors in different input regions indicate the failure of stationarity assumption used for emulator construction. Then, we can fit the $\lambda_l(\stanx)$ via, for example, categorical regression 
 \[ \lambda_l(\stanx) =  \frac{\exp(g(\bb{x})^T \bb{\alpha}_l)}{\sum_{\dashed{l} = 1}^L \exp(g(\bb{x})^T \bb{\alpha}_{\dashed{l}})}, \]
 with parameters $A = (\bb{\alpha}_1, \dots, \bb{\alpha}_L)^T$ and $\bb{\zeta}=(\zeta_1, \dots, \zeta_L)$ and suitable prior $\pi(A, \bb{\zeta})$.

The $\bb{\lambda}(\stanx)$ are computed for $M$ posterior samples (after warm-up) and fixed at the mean value over the posterior samples denoted by $\hat{\bb{\lambda}}(\stanx) = (\hat{\lambda}_1(\stanx), \cdots, \hat{\lambda}_L(\stanx))$:
 \begin{equation*}
 \hat{\bb{\lambda}}(\stanx) = \frac{1}{M}\sum_{m=1}^M \bb{\lambda}(\stanx; A_m).
 \end{equation*} 
 Our covariance function from \cref{eq:covarnonstat} becomes:
\begin{equation*}
k(\stanx, \stanx'; \bb{\sigma}^2, \Delta, \bb{\tau}^2) = \sum_{l=1}^L\hat{\lambda}_l(\stanx)\hat{\lambda}_l(\stanx')k_l(\stanx, \stanx'; \sigma_l^2, \bb{\delta}_l) + \mathbb{1}\big\{ \stanx = \stanx'\big\} \sum_{l=1}^L z_l(\stanx) z_l(\stanx')\tau_l^2,
\end{equation*}
where
\begin{equation*}
z_l(\stanx) = \begin{cases}
1, &\hat{\lambda}_l(\stanx) = \max_{k}\big\{\hat{\lambda}_k(\stanx) \big\},\\ 
0, &\mbox{otherwise}.
\end{cases}
\end{equation*}  
Fixing $\bb{\lambda}(\stanx)$ at $\hat{\bb{\lambda}}(\stanx)$ in this way resembles the common and effective Cross Validation (CV) approach to estimate the parameters of the statistical model, i.e. the model parameter values that provide the user with the smallest LOO error are chosen. For example, \cite{Bachoc2013} uses the LOO-CV to estimate $\sigma^2$ and $\bb{\delta}$ for a GP model and numerically proves that the Cross-Validation (CV) estimation is more robust than the Maximum Likelihood estimation (ML) of parameters to model misspecification, the case when the true underlying covariance function does not belong to the family of covariance functions that they are operating with.  

Interestingly, \cite{Ba2012} also consider the squared residuals $\bb{s}^2$ for a global trend as a measure of local simulator behaviour in a similar manner to how we consider standardized residuals after fitting the stationary GP model.  The squared residuals appears in the model of the standardized volatility function $v(\stanx)$:
\begin{equation*}
v(\stanx) = \frac{\bb{g}_b^T(\stanx)\bb{s}^2}{\bb{g}_b^T(\stanx)\bb{1}}
\end{equation*}
where $\bb{g}_b^T(\stanx)=(g_b(\stanx - \stanx_1), \cdots, g_b(\stanx - \stanx_n))$ is the global trend correlation function with extra parameter $b$ evaluated between the point of interest $\stanx$ and the design points. For further details refer to \cite{Ba2012}.

Finally, the optimal choice of $L$ is determined by considering WAIC (widely applicable information criteria) \cite{Watanabe2010} for mixture models with $L=1, \dots, 4$ and choosing the model that provides the lowest WAIC and/or the biggest improvement in the WAIC. In application, if $L=4$ was selected, we would check higher values of $L$ to ensure that we hadn't missed any complex structures. WAIC estimates the pointwise-out-of-sample prediction accuracy from a fitted model using log-likelihood evaluated at the draws from the posterior distribution of parameters. Contrary to BIC  \cite{Schwarz1978}, WAIC uses the entire posterior distribution and it is asymptotically equal to Bayesian cross-validation \cite{Gelman2014}. In a similar manner \cite{Fuentes2001} considers the improvement in AIC (Akaike information criterion) \cite{Akaike1974} and \cite{Banerjee2004} consider the improvement in BIC score \cite{Schwarz1978} with the increase in $L$ for their spatial models.
\subsection{Priors for model hyperparameters}
\label{sec:Priors}
For the mixture model we specify $g(\stanx)=(\mathrm{x}_1, \dots, \mathrm{x}_p)^T$ and priors for parameters, $\zeta_l \sim \log N(-1, 1)$ and $\bb{\alpha}_l\sim \text{Normal}(0, 5)$, with $l=1, \dots, L$ and $\bb{\alpha}_l=(\alpha_{1l}, \dots, \alpha_{pl})$, the probability distributions with wide support, representing weak prior information \cite{Almond2014, StanDevelopmentTeam2017}. We constrain the prior standard deviation parameter to follow the ordering, $\zeta_1\leq \zeta_2 \leq \dots \leq \zeta_L$, solving the problem of having multiple modes in the posterior distribution for mixture models, and ensuring good mixing of our Markov chains \cite{Almond2014}.

We use Stan \cite{StanDevelopmentTeam2017}, based on Hamiltonian Monte Carlo (HMC), for our inference, to enable users of our code to be very flexible with their prior choices. HMC does not provide sampling for discrete parameters \cite{StanDevelopmentTeam2017}, therefore, the posterior of the discrete group allocation indices, $s=(s(\stanx_1), \dots, s(\stanx_n))$, cannot be sampled directly and so we integrate $s$ out in the likelihood. Mixture components are computed a posteriori. The regression coefficient parameters are sampled from the joint posterior, integrating over $s$, 
\[ p(A\vert \bb{e}, \bb{\zeta}) \propto \int p(\bb{e} \vert s, \bb{\zeta}) p(s\vert A)p(A)p(\bb{\zeta}) ds.\]
Sinse $s$ is discrete, this is equivalent to 
\[ p(A\vert \bb{e}, \bb{\zeta}) \propto \prod_{i=1}^n\Bigg( \sum_{l=1}^L Pr(s(\stanx_i)=l \vert A) p(e_i \vert \zeta_l)\Bigg) p(A) p(\bb{\zeta}) \]
where $A=(\bb{\alpha}_1, \dots, \bb{\alpha}_L)^T$ and $\bb{\zeta}=(\zeta_1, \dots, \zeta_L)$.

\begin{figure}[ht]
\begin{center}
\includegraphics[width=0.85\textwidth,height=0.2\textheight]{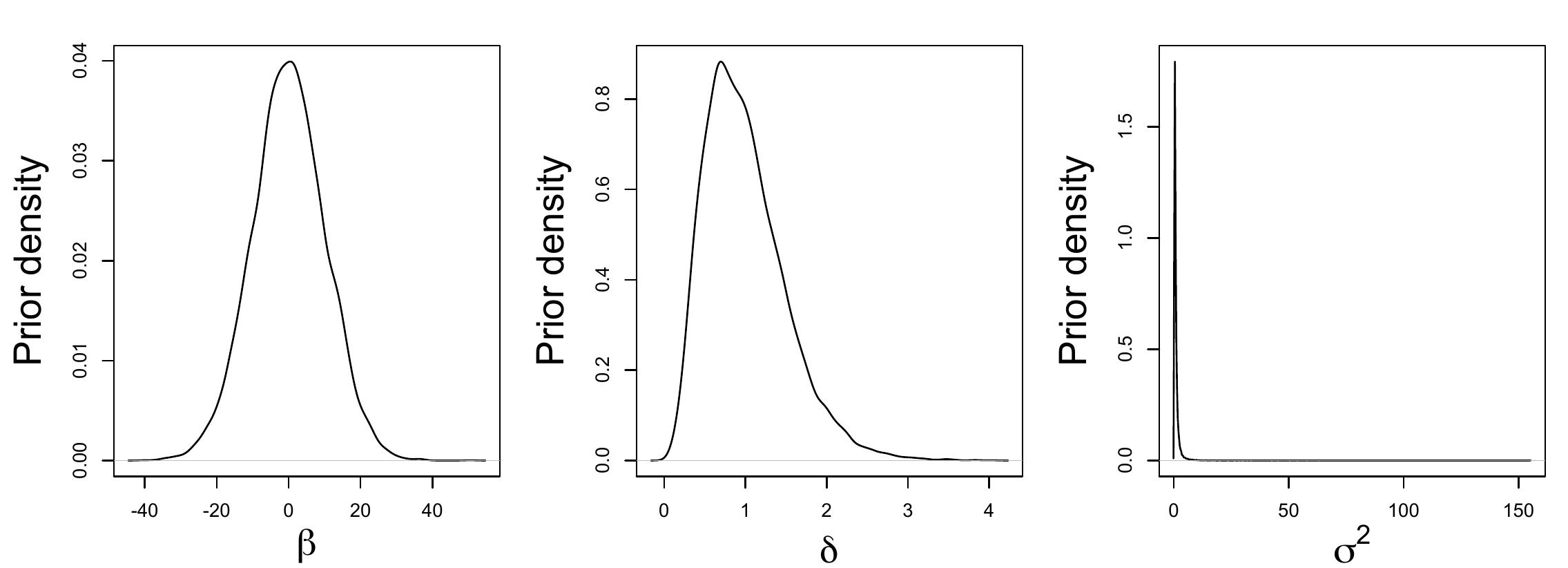}
\end{center}
\caption{Density plots of Normal(0, 10) (\textit{left}), Gamma(4, 4) (\textit{center}) and InvGamma(2, 1) (\textit{right}) distributions.}
\label{fig:figureprior}
\end{figure}

For GP emulator we specify a linear structure for the prior mean,  $h(\stanx) = (1, x_1, \dots, x_p)^T$ and weakly informative prior, $\bb{\beta}_{1:(p+1)} \sim N(0, 10)$. It is crucial to formulate priors for $\bb{\delta} = (\delta_1, \dots, \delta_p)$ and $\sigma^2$ as the numerical issues could arise in Stan from estimating GP model with improper form of priors for these parameters \cite{StanDevelopmentTeam2017}. The weakly informative prior $\delta_k\sim\text{Gamma}(4, 4)$, $k=1, \dots, p$, is used for correlation length parameters. From \cref{fig:figureprior} it could be seen that this form of prior restricts the extremely small correlation length as well as constrains the large correlation length values. Small correlation length value would lead to the flat likelihood and the model overfitting to the input data, whereas large correlation length value leads to the linear posterior with respect to a particular input variable, i.e. no effect from this input on the final estimated GP model. \cite{StanDevelopmentTeam2017}. 
We define a weakly informative prior, $\sigma^2\sim IG(2, 1)$, commonly used in the literature \cite{Gramacy2008, Montagna2016}. Finally, we also assign $\tau^2$ to an arbitrary small value, i.e. $\tau^2=0.0001$, to facilitate stable matrix inversion as suggested by \cite{Andrianakis2012}. We specify the same forms of prior distribution for model parameters $\bb{\beta}$ and $\big\{\sigma_l, \bb{\delta}_l \big\}_{l=1:L}$ for our nonstationary GP emulator.

\section{Case studies}
\label{sec:case}
We consider the performance of stationary and our proposed nonstationary GP emulators on a set of test functions and also demonstrate the performance of Bayesian TGP \cite{Gramacy2008} and composite GP (CGP) \cite{Ba2012} for comparison. To implement TGP and CGP, we used \textsf{R} packages \textsf{TGP} with default settings and fixed nugget parameter at 0.01, note the nugget has a different interpretation to $\tau^2$ in TGP \cite{Gramacy2007},  and \textsf{CGP} with default settings \cite{Ba2018}. Both \textsf{R} packages are available from \textsf{CRAN}. 

We assess the performance of the mentioned above GP emulators by considering Root Mean Squared Error (RMSE) and Interval Score for $(1-\alpha)\times 100\%$ prediction interval \cite{Gneiting2007}. The interval score with lower $l$ and upper $u$ endpoints at level $\frac{\alpha}{2}$ and $1-\frac{\alpha}{2}$ quantiles is found
\begin{equation*}
S_{\alpha}^{int}(l, u; y) = (u-l) + \frac{2}{\alpha}(l-y)\mathbb{1}\big\{ y<l\big\} + \frac{2}{\alpha}(y-u)\mathbb{1}\big\{ y>u\big\}.
\end{equation*}
We are interested in the low values of the Interval Score by being rewarded for the narrow prediction interval and being penalized if $y$ misses the prediction interval. The size of the penalty depends on $\alpha$ and we specify $\alpha=0.05$.

Prior to fitting our emulators we perform the transformation on ensemble, $\big\{\bb{X}, \bb{F} \big\}$, i.e. we transform $\bb{X}$ on $[-1, 1]$ scale and center the corresponding response, $\bb{F}$, around zero by subtracting mean of $\bb{F}$ and scaling by its standard deviation. 

\subsection{The 2D `wavy' function}
\label{subsec:Wavy}
\begin{figure}[ht]
\begin{center}
\includegraphics[width=0.5\textwidth,height=0.35\textheight]{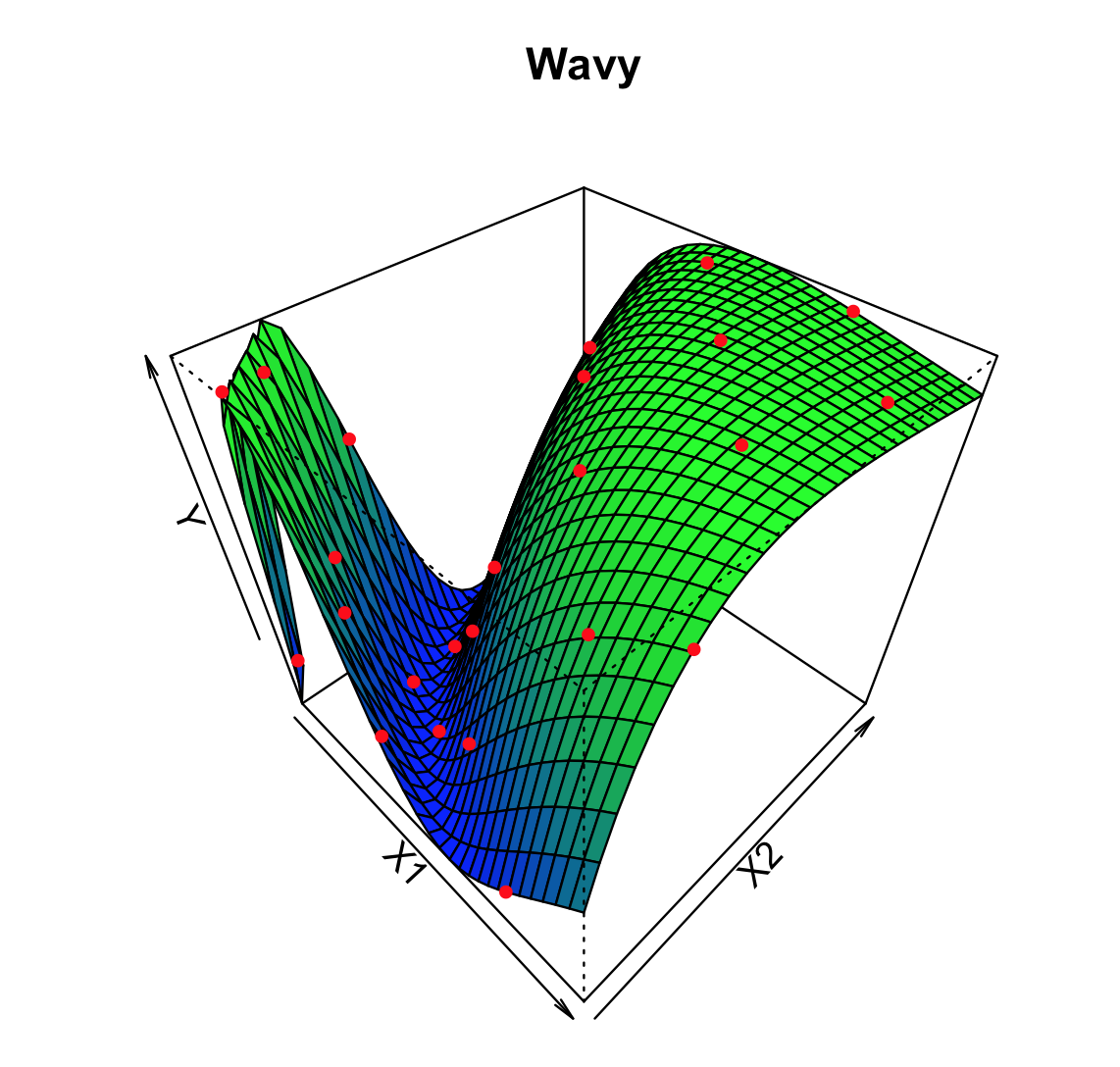}
\end{center}
\caption{True function for the two-dimensional numerical example and 24-run maximin distance LHD red points used as design.}
\label{fig:figure2}
\end{figure}

In this subsection, we examine the performance of emulators on the nonstationary function considered by \cite{Ba2012}. The `wavy' function has the equation $f(\mathrm{x}_1, \mathrm{x}_2)=\sin(1/(0.7\mathrm{x}_1+0.3)(0.7\mathrm{x}_2+0.3))$ $(\mathrm{x}_1, \mathrm{x}_2 \in [0, 1])$. It fluctuates rapidly when $\mathrm{x}_1$ and $\mathrm{x}_2$ is small, but gradually becomes smooth as $\mathrm{x}_1$ and $\mathrm{x}_2$ increases toward 1. See  \cref{fig:figure2}.

We use a 24-run maximin distance Latin Hypercube (LHC) \cite{Morris1995} to train our emulators. Firstly, we construct a stationary GP emulator and consider the standardized errors by fitting the mixture model with $L=1, \dots, 4$.

\begin{figure}[ht]
\begin{center}
\includegraphics[width=0.7\textwidth, height=0.2\textheight]{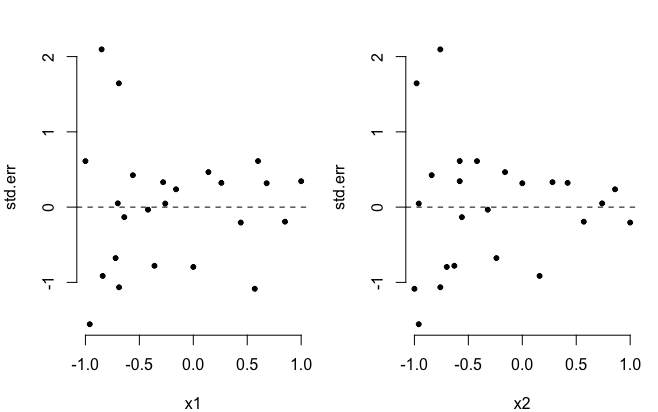}
\includegraphics[width=0.7\textwidth, height=0.2\textheight]{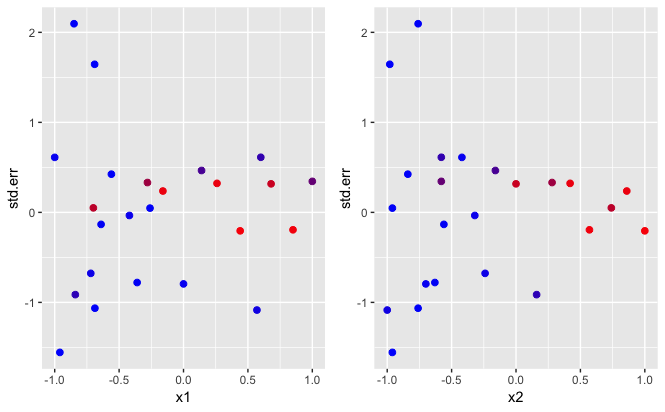}
\end{center}
\caption{\textit{Top row}: $e_i$ against $\mathrm{x}_1$ and $\mathrm{x}_2$. \textit{Bottom row}: coloured $e_i$: the deep blue colour corresponds to the higher probability of a point being allocated to region 1, while the deep red colour corresponds to the higher probability of a point being allocated to region 2.}
\label{fig:figure3}
\end{figure}

\begin{table}[h!] 
\caption{Mixture Model Comparison for $L=1, 2, 3, 4$.}\label{table:1}
\centering
\begin{tabular}{|c|c|c|c|}
\hline
Models & WAIC & Models & $\Delta$ WAIC\\ [0.5ex]
\hline
$L=1$ & 60.84 & $(L=1)$ - $(L=2)$ & 7.52\\
\hline
$L=2$ & 53.32 & $(L=2)$ - $(L=3)$& -0.54\\
\hline
$L=3$ & 53.86 & $(L=3)$ - $(L=4)$& -0.56 \\
\hline
$L=4$ & 54.42& & \\ [0.5ex]
\hline
\end{tabular}
\label{tab:table1}
\end{table}

\cref{tab:table1} demonstrates that the mixture model with $L=2$ has the lowest WAIC measure as well as the largest improvement in WAIC and we conclude that $L=2$ is the optimal number of the input regions. Interestingly, TGP model partitions the input space into two separate regions as well.

\cref{fig:figure3} confirms that $L=2$ is a good choice for the mixture model, i.e.  the variability of the errors depends on both inputs and there are two distinct identifiable regions of standardized error behaviour. For $\mathrm{x}_1, \mathrm{x}_2 <-0.5$, the standardized errors exhibit large variability, i.e. standardized errors ranging from -1.5 to 2, while for input values greater than -0.5 the standardized errors variability significantly decreases. We use the model described in \cref{sec:Mixture} to derive mixing functions. 

\begin{figure}[ht]
	\begin{center}
		\includegraphics[height=0.3\textheight, width=1\textwidth]{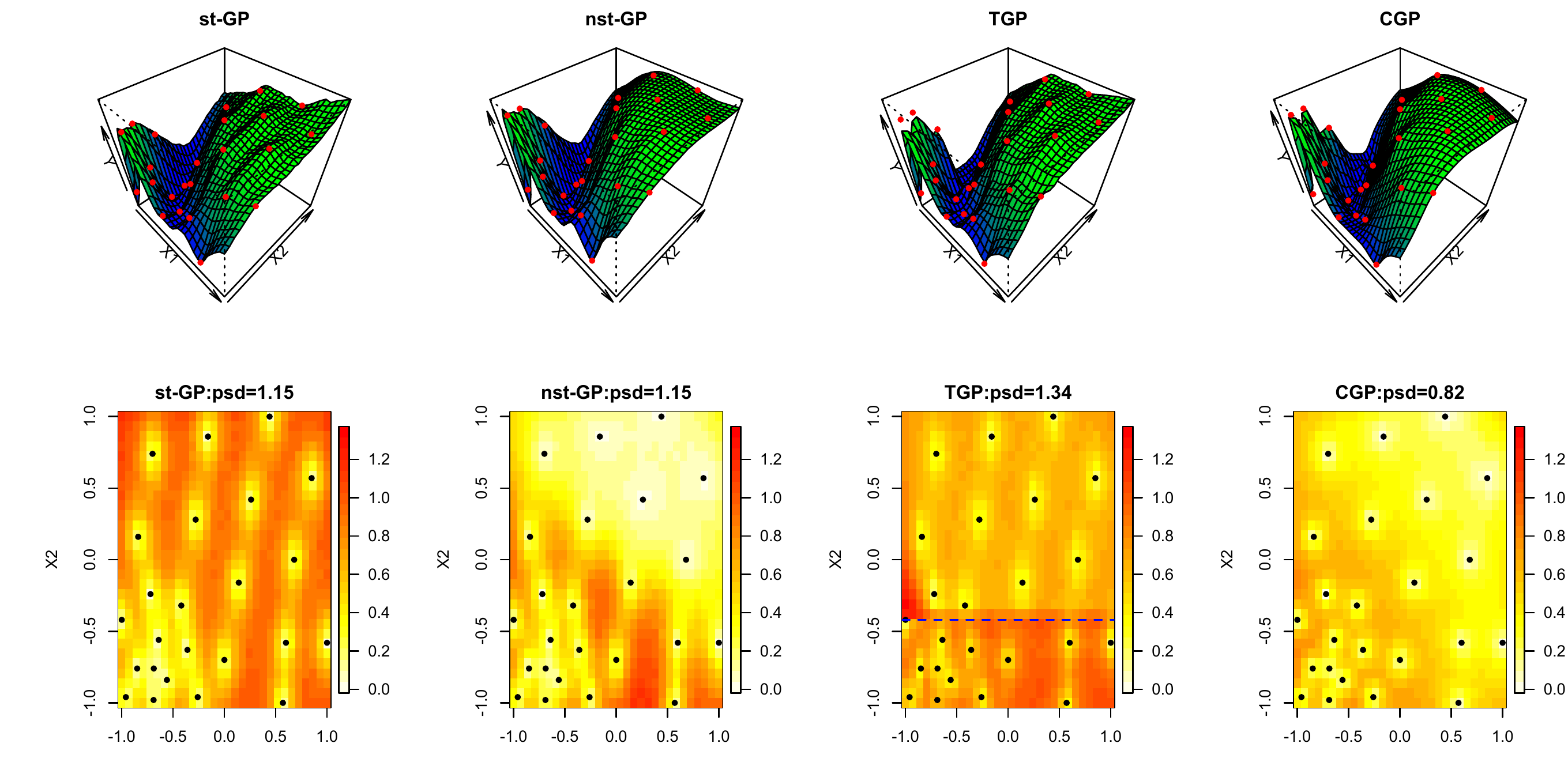}
		\includegraphics[height=0.15\textheight, width=1\textwidth]{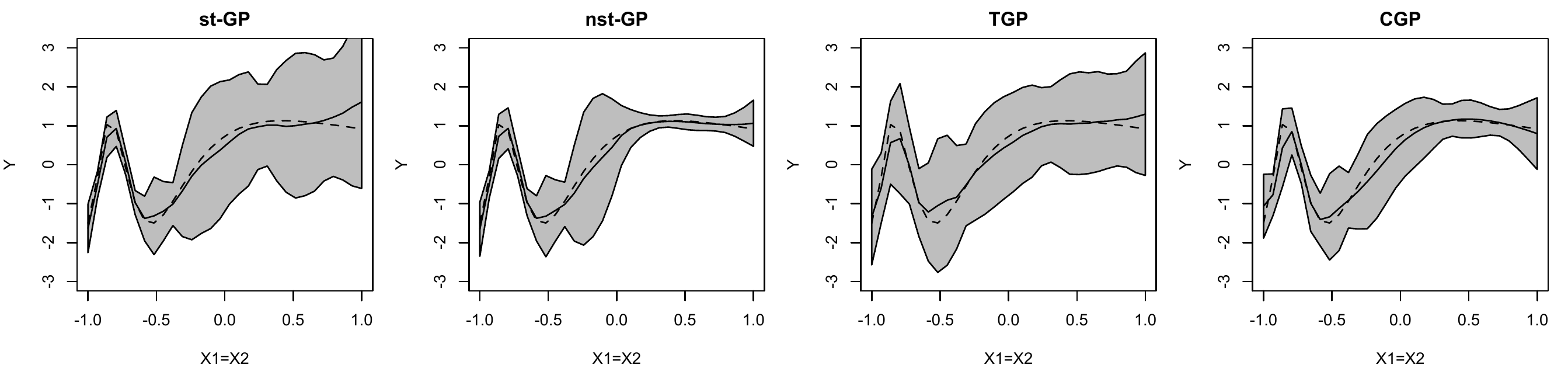}
	\end{center}
	\caption{Comparison between stationary GP (\textit{st-GP}), our nonstationary GP (\textit{nst-GP}), TGP, and CGP. \textit{First row}: posterior predictive mean surface. \textit{Second row}: posterior predictive standard deviation and maximum posterior predictive standard deviation (max psd). \textit{Third row}: cross-section performance at $\mathrm{x}_1=\mathrm{x}_2$. The dashed line is the true function, the solid black line is the posterior mean predictive curve, and the grey areas denote two standard deviation prediction intervals.}
	\label{fig:figure4}
\end{figure}

\cref{fig:figure4} demonstrates the performance of all four emulators for the toy function. The predictive mean surface, $m^{*}(\stanx)$, (top row) produced by our nonstationary emulator and by CGP resemble the image of the true function surface, while the stationary GP emulator and TGP emulator produce a number of ridges across the input space.

From the second row of \cref{fig:figure4} we observe that the stationary GP emulator produces the lowest values of predictive standard deviation around the design points, but due to the small values of the correlation length parameters, the information is quickly `dying' away from the design points. Our nonstationary GP emulator and CGP manage to learn and explore the function behaviour for high values of $\mathrm{x}_1$ and $\mathrm{x}_2$ based on a few design points due to the stronger correlation structure in this region. TGP partitions the input space at $\mathrm{x}_2=-0.42$ (blue dashed line) and we observe the highest predictive standard deviation around the partition.

The behaviour of the wavy function changes along the line where $\mathrm{x}_1 = \mathrm{x}_2$. \cref{fig:figure4} demonstrates the cross-section plot for the stationary model and shows the limitations of the stationary emulator, i.e. it is overconfident in the region where the function behaviour changes rapidly. Both the TGP and CGP models are under-confident across the cross-section, while our nonstationary GP emulator produces the lowest prediction intervals among all of the methods, through is perhaps under-confident during the transition to smoother behaviour $\mathrm{x}_1, \mathrm{x}_2 \geq- 0.5$.

\begin{table}[h!] 
\caption{Interval Score (IS) and Root Mean Squared Error (RMSE) for the 2D `wavy' function}
\centering
\begin{tabular}{|c|c|c|}
\hline
Models & IS & RMSE\\ [0.5ex]
\hline 
st-GP & 2.76 & 0.298 \\
\hline
nst-GP & 1.52 & 0.278 \\
\hline
TGP & 2.67 & 0.267\\
\hline
CGP & 1.61 & 0.233\\ [0.5ex]
\hline
\end{tabular}
\label{tab:table2}
\end{table}

\cref{tab:table2} demonstrates that our nonstationary GP emulator produces the lowest Interval Score for the validation ensemble followed by CGP. However, CGP and TGP demonstrate lower RMSE than our proposed nonstationary GP emulator.

\subsection{Nonstationarity in 5 dimensions}
\label{subsec:FiveD}

\begin{figure}[ht]
	\begin{center}
		\includegraphics[height = 0.25\textheight, width=0.7\textwidth]{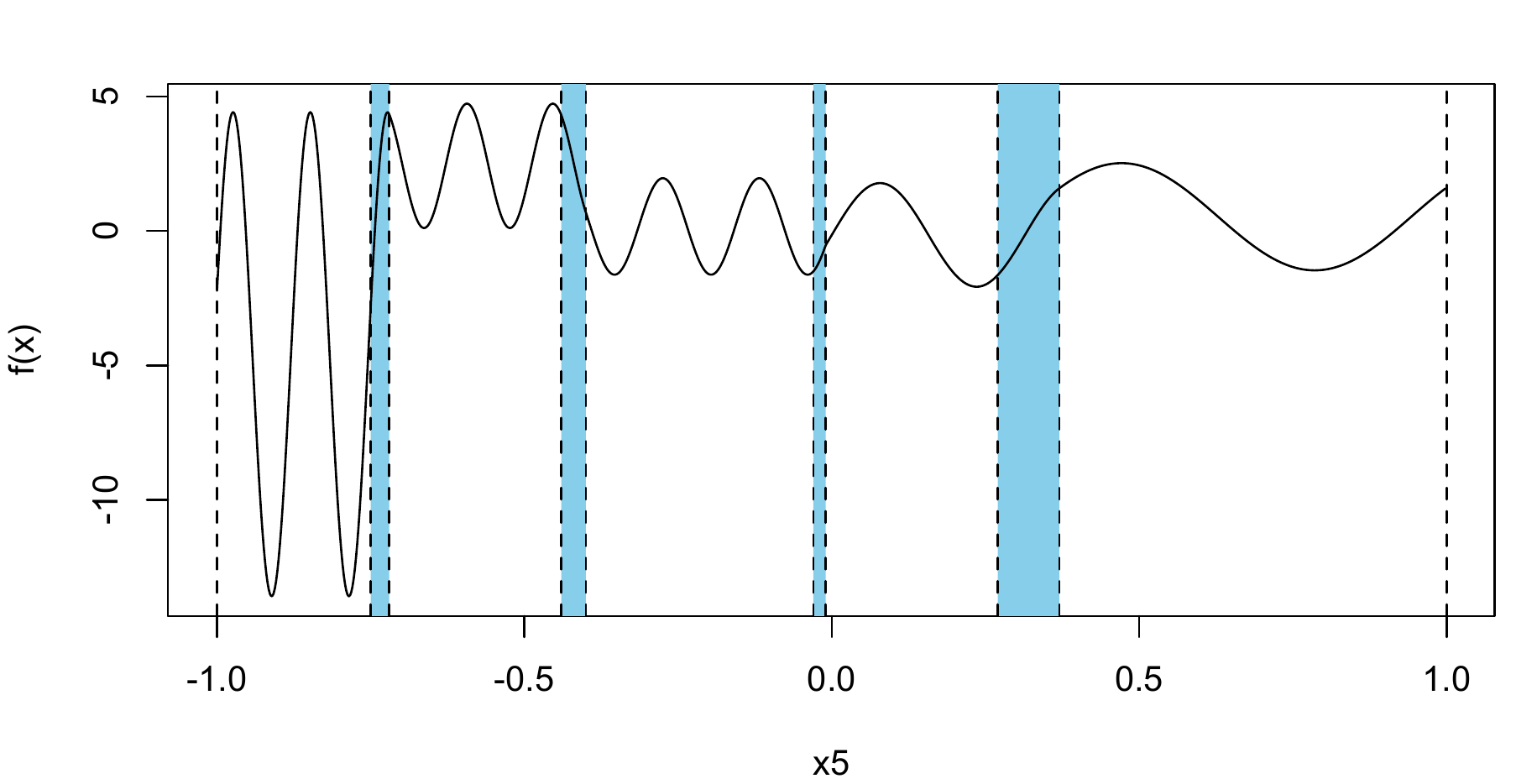}
	\end{center}
	\caption{Our 5D function $f(\stanx)$ plotted against $\mathrm{x}_5$}
	\label{fig:figure5}
\end{figure}

We now examine the performance of our method in higher dimensions, using the 5D function
\begin{equation}
 f(\stanx) = \beta_0+\beta_1\mathrm{x}_1 + \beta_2\mathrm{x}_2+\beta_3\mathrm{x}_3+\beta_4\mathrm{x}_4+\beta_5\sin(\beta_6\mathrm{x}_5), \nonumber
\end{equation}
with $\beta_5, \beta_6$ changing in each of 5 different intervals in $\mathrm{x}_5$, as shown in \cref{fig:figure5}. Intervals in white correspond to five separate function behaviours, while the intervals in blue are a mixture of functions from two neighbouring regions to ensure continuity. Our choices of $\beta_5, \beta_6$ impose significant variability in smoothness with changes in $\mathrm{x}_5$. We choose to vary the stationarity properties along one axis in order to favour TGP, which partitions the input space along one input. 

We generate a 4-extended LHC of size 100 following the methodology presented by \cite{Williamson2015}. The 100 member LHC is composed of four, 25 member LHCs, each added sequentially, ensuring that the composite design is orthogonal and space-filling at each stage of extension. We will compare the performance of all methods using Leave One Latin Hypercube Out (LOLHO) diagnostics, i.e. each row in \cref{fig:figure8} represents the predictions generated for a left out from the design LHC by refitted emulators. LOLHO diagnostics offer a sterner test for an emulator than LOO diagnostics and allow us to assess which areas of the input space do not validate well \cite{Williamson2015}.

\begin{figure}[ht]
	\begin{center}
		\includegraphics[height = 0.4\textheight, width=0.8\textwidth]{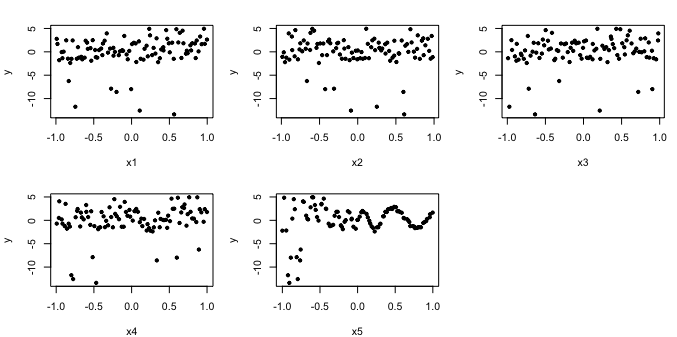}
	\end{center}
	\caption{The response of 5D function $f(\stanx)$ against all five inputs.}
	\label{fig:figure6}
\end{figure}

Before constructing our emulators we produce the scatter plots of function response against the inputs. From \cref{fig:figure6} we observe the variability in the function response  $f(\stanx)$ is mainly driven by $\mathrm{x}_5$. The input $\mathrm{x}_5$ could be considered as an `active input' and the global trend together with the residual term could be modelled in terms of the `active inputs' with the nugget term to account for the remaining variation \cite{Craig2001, Goldstein2009, Cumming2009}. However, operating within a Bayesian framework and with flexible priors we could specify stronger prior information for $\delta_5$ \cite{Higdon2008, Williamson2014} for our stationary and nonstationary GP emulators. We achieve this by keeping the same $\text{Gamma}(4, 4)$ prior for $\delta_5$ but specifying a smoother prior in the other 4 dimensions via $\delta_1, \dots, \delta_4 \sim \text{Gamma}(42, 9)$ (this distribution was chosen by using the MATCH elicitation tool \cite{Morris2014} to capture a reasonable distribution giving more weight to longer correlation lengths).

\begin{figure}[ht]
\begin{center}
\includegraphics[height = 0.15\textheight, width=1\textwidth]{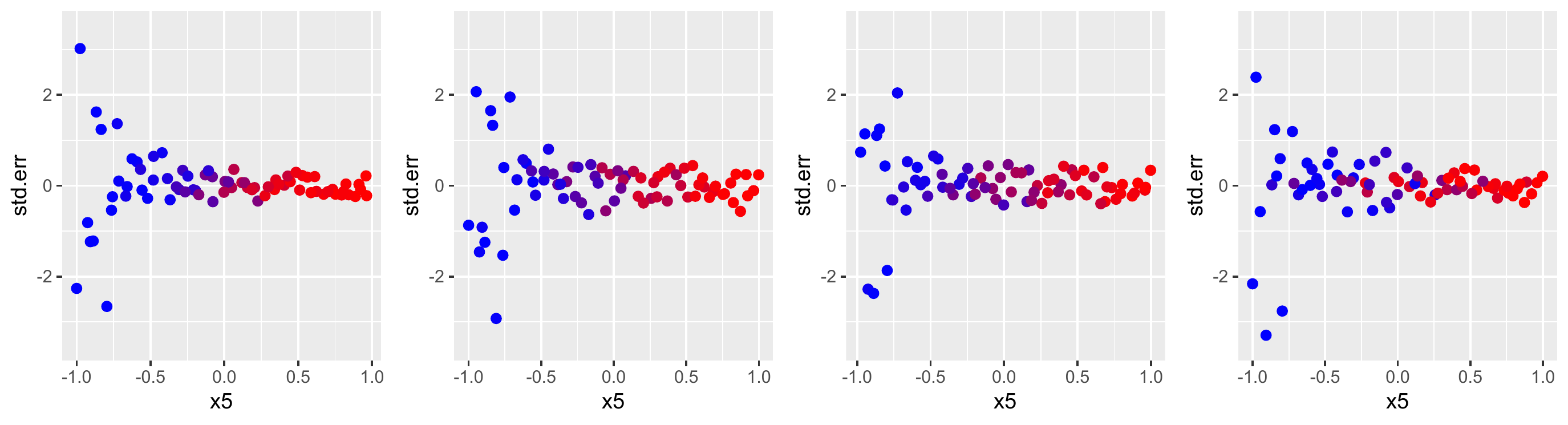}
\end{center}
 \caption{$e_i$ against the $\mathrm{x}_5$ for four sub-designs. The deep blue colour corresponds to the higher probability of a point being allocated to region 1, while the deep red colour corresponds to the higher probability of a point being allocated to region 2. }
\label{fig:figure7}
\end{figure}

\begin{figure}[ht]
\begin{center}
\includegraphics[height = 0.55\textheight, width=1\textwidth]{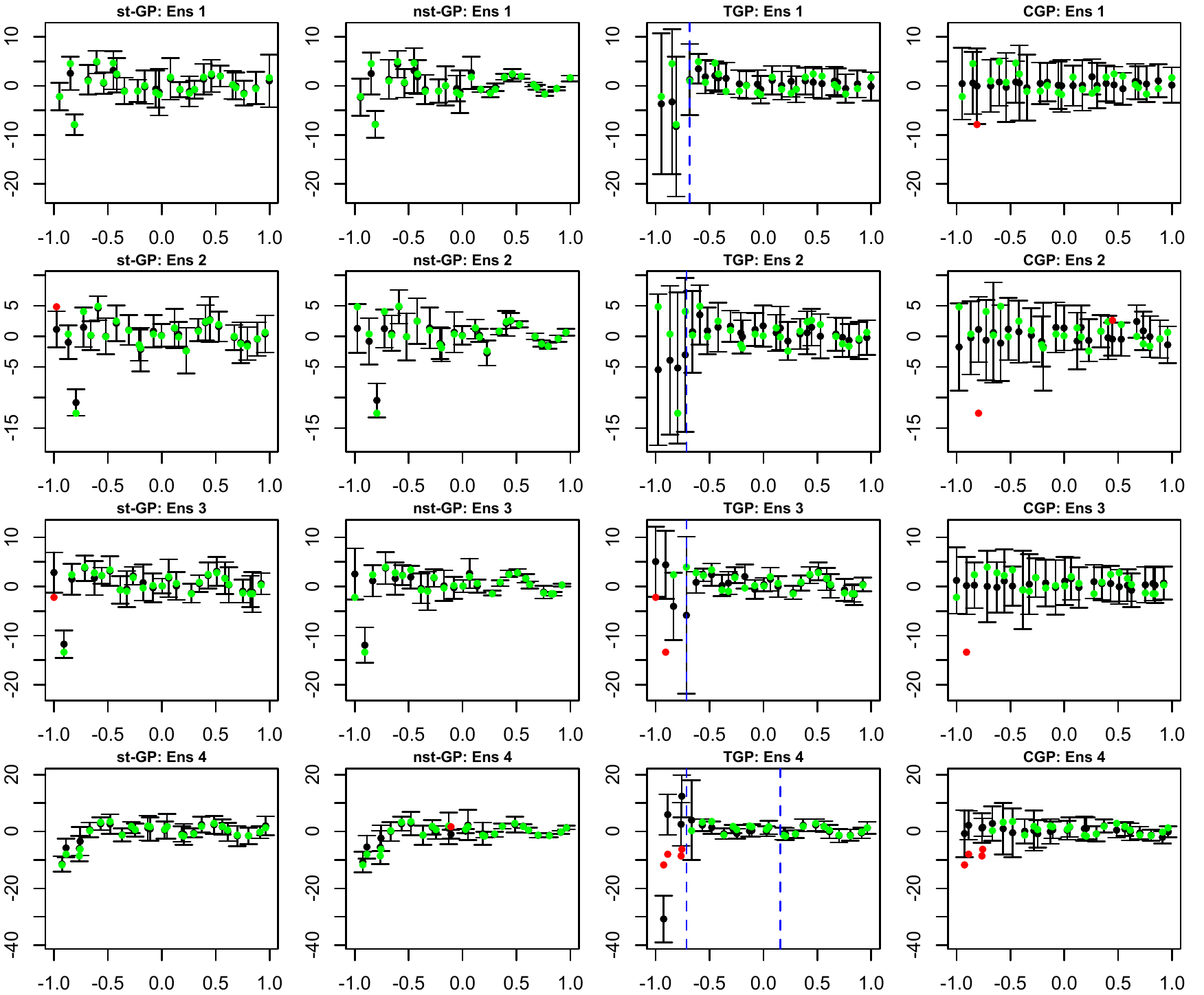}
\end{center}
\caption{Leave One Latin Hypercube Out (LOLHO) plots for stationary GP (\textit{st-GP}), our non-stationary GP (\textit{nst-GP}), TGP and CGP. Each row is constructed by leaving one LHC out. Blue dashed lines correspond to the partitions produced by TGP. The posterior mean and two standard deviation prediction intervals produced by emulators are in black. The true function values are in green if they lie within two standard deviation prediction intervals, or red otherwise.}
\label{fig:figure8}
\end{figure}

To construct our nonstationary GP emulator we consider the mixture model for standardized errors for each left out LHC derived from a stationary fit with $L=1, \dots, 4$. We observe from \cref{tab:table5D} that the mixture model with $L=2$ offers the largest improvement in WAIC for all four ensembles, despite the fact that the true function has $L=5$ in practice. \cref{fig:figure7} demonstrates the performance of mixture model with $L=2$ for all four ensembles. 

\begin{table}[h!] 
\caption{Mixture Model Comparison for $L=1, 2, 3, 4$ for Ensembles. The first four rows of the table correspond to WAIC and the last three rows correspond to the improvement in WAIC.}
\centering
\begin{tabular}{|c|c|c|c|c|}
\hline
Models & Ens 1 & Ens 2 & Ens 3 & Ens 4\\ [0.5ex]
\hline 
$L=1$ &162.04 & 160.88 & 145.13 & 168.24\\
\hline
$L=2$ & 81.42 & 115.33 & 101.45 & 94.80\\
\hline
$L=3$ & 58.79 & 102.67 & 75.50 & 78.15\\
\hline
$L=4$ & 58.62 & 100.12 & 76.34 &  74.45\\
\hline
$(L=1) - (L=2)$  & 80.62 & 45.55 & 43.68 & 73.43\\ 
\hline
$(L=2) - (L=3)$  & 22.63 & 12.66 & 25.95 & 16.65\\
\hline
 $(L=3) - (L=4)$  & 0.17 & 2.55 & -0.84 & 3.70\\ [0.5ex]
 \hline
\end{tabular}
\label{tab:table5D}
\end{table}

\begin{table}[h!] 
\caption{Interval Score for the 5D function example.}
\centering
\begin{tabular}{|c|c|c|c|c|}
\hline
Models & Ens 1 & Ens 2 & Ens 3 & Ens 4\\ [0.5ex]
\hline 
st-GP &6.10 & 6.63 & 7.25 & 6.22\\
\hline
nst-GP & 4.40 & 4.16 & 3.92 & 6.64\\
\hline
TGP & 8.99 & 9.62 & 23.84 & 58.66\\
\hline
CGP &  9.92 & 22.91 & 21.69 & 35.80\\ [0.5ex]
\hline
\end{tabular}
\label{tab:table4}
\end{table}

\begin{table}[h!] 
\caption{Root mean squared error (RMSE) for the 5D function example.}
\centering
\begin{tabular}{|c|c|c|c|c|}
\hline
Models & Ens 1 & Ens 2 & Ens 3 & Ens 4\\[0.5ex]
\hline 
st-GP &0.580 & 1.028 & 1.152 & 0.993\\
\hline
nst-GP & 0.619 & 1.042 & 1.117 & 1.312\\
\hline
TGP & 2.068 & 3.229 & 4.574 & 6.511\\
\hline
CGP &  2.606 & 3.728 & 3.307 & 4.227\\[0.5ex]
\hline
\end{tabular}
\label{tab:table5}
\end{table}

\cref{fig:figure8} shows the LOLHO diagnostic plots for each of our 4 emulators (columns) for each of the sub-designs considered (rows). We observe that the main issue with our stationary GP emulator is the same length of prediction intervals across the whole range of $\mathrm{x}_5$ values, i.e. it fails to recognise the changes in the function response variability. TGP partitions the input space into two separate regions and constructs the Gaussian process model in each region individually. Despite TGP performing relatively well in the region where the function is `well-behaved', it performs poorly for small values of $\mathrm{x}_5$ across all ensembles, perhaps due to a small number of design points in this region. CGP does recognise the variability of function response, however it is overconfident for all four ensembles. In particular both methods do not perform well for the Ensemble 4, which requires further investigation. Our nonstationary GP emulator with $L=2$ performs well and produces smaller prediction intervals in the region when the function is `well-behaved', i.e. for $\mathrm{x}_5$ close to 1. For ensemble 2 and ensemble 4 we produce 2 failures (red dots), which is consistent with our uncertainty specification \cite{Williamson2015}. \cref{tab:table4} and \cref{tab:table5} demonstrate that our nonstationary GP emulator produces the lowest Interval Score for all sub-designs and Root Mean Squared Error values for the majority of ensembles. 

\section{Experiments with ARPEGE-Climat model}
\label{sec:cnrm}
In this section we present work that has been done as part of the ANR (Agence Nationale de la Recherche) funded HIGH-TUNE project. The primary objective of the project is to improve and tune the boundary-layer cloud parameterizations of two French General Circulation Models, ARPEGE-Climat and LMDZ. LMDZ is developed at the Laboratoire de M\'{e}t\'{e}orologie Dynamique (LMD) and is the atmospheric component of the IPSL (Institute Pierre Simon Laplace) climate model. ARPEGE-Climat is developed at the Centre National de Recherches M\'{e}t\'{e}orologiques (CNRM) and is the atmospheric component of the CNRM climate model. In the present study, simulations with the version 6.3 of ARPEGE-Climat are used. This is an updated version compared to the one described in \cite{Voldoire2013} (see also \cite{Abdel2018} for further details).

Boundary-layer clouds are crucial components of the water and energy budget of the climate system \cite{Bony2005}. These clouds are not explicitly resolved by solvers at the resolution used for running GCM's. To account for their impact on the resolved scale, they are parameterized through a set of approximating equations, dependent on a number of `free' (internal) parameters.  In the climate community the estimation of these parameters is called tuning \cite{Hourdin2017}. 

\begin{figure}[ht]
\begin{center}
\includegraphics[height=0.5\textheight, width=0.75\textwidth]{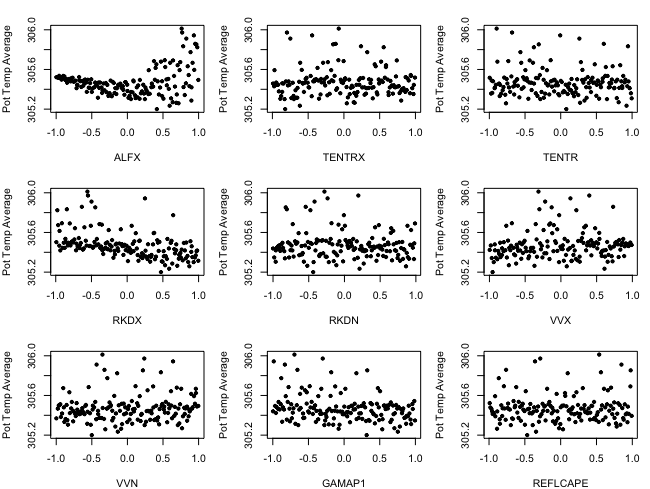}
\end{center}
\caption{Average potential temperature against nine standardized inputs to SCM (Single Column Model).}
\label{fig:figure9}
\end{figure}

The HIGH-TUNE project aims to improve the representation of boundary-layer clouds by comparing single-column simulations of the GCM's (SCM's) over a variety of cases with reference 3D high-resolution Large-Eddy simulations (LES) run with exactly the same setup. History matching \cite{Williamson2015a, Williamson2017}, a statistical approach to tuning, is then performed in order to rule out part of the free parameter space, where the SCM's are not effectively mimicking the LES results, and provide guidance to the GCM developers in the choice of the values of these parameters.
It is a common practice to treat LES as surrogate observations for parameterization development and tuning, because it is difficult to observe real clouds at the required temporal and spatial scales \cite{Hourdin2013}. In climate modelling this concept is termed `process-based tuning'. 

Emulators for the SCM's will be required whenever new parameterizations are developed and tested. In our work with the HIGH-TUNE team we have found using routine stationary GP's insufficient for this task and so we present the performance of our nonstationary method within this application. The tuning, what we would term calibration, of these models is beyond the scope of this paper. We will consider the average potential temperature generated by the SCM by varying nine input parameters of interest, all associated with the parameterization of convection.

Firstly we identified the physically plausible ranges of the input model parameters with the HIGH-TUNE team and standardized these to the range [-1, 1], which is the usual practice in constructing emulators for computer experiments. We generated a 160 member LHC composed of four, 40 member LHCs \cite{Williamson2015}, similar to the design for the 5D example in \cref{subsec:FiveD}.

\cref{fig:figure9} plots the average potential temperature against each input. We see that the average potential temperature varies most with the input ALFX, i.e. the variability in the response of the average potential temperature increases as ALFX increases.

\begin{figure}[ht]
\begin{center}
\includegraphics[height=0.15\textheight, width=1\textwidth]{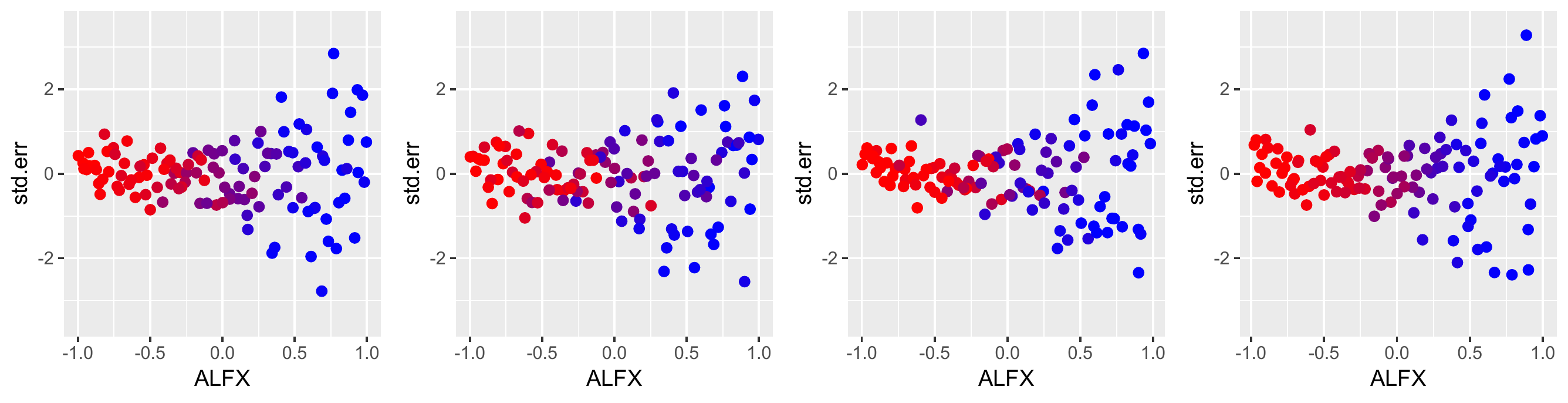}
\end{center}
\caption{Coloured $e_i$ against ALFX for four sub-designs. The deep blue colour corresponds to the higher probability of a point being allocated to region 1, while the deep red colour corresponds to the higher probability of a point being allocated to region 2.}
\label{fig:figure10}
\end{figure}

\begin{table}[h!] 
\caption{Mixture Model Comparison for $L=1, 2, 3, 4$ for Ensembles. The first four rows of the table correspond to WAIC and the last three rows correspond to the improvement in WAIC.}
\centering
\begin{tabular}{|c|c|c|c|c|}
\hline
Models & Ens 1 & Ens 2 & Ens 3 & Ens 4\\ [0.5ex]
\hline
$L=1$ & 295.15 & 299.51& 297.85& 303.19\\
\hline
$L=2$ & 254.85 & 270.59 & 245.38 & 262.10\\
\hline
$L=3$ & 246.14 & 266.31 & 245.77 & 250.95\\
\hline
$L=4$ &  245.54 & 268.10 & 247.88 & 251.94\\
\hline
$(L=1) - (L=2)$  & 40.30 & 28.92 & 52.47 & 41.09\\ 
\hline
$(L=2) - (L=3)$  & 8.71 & 4.28 & -0.39 & 11.15\\
\hline
$(L=3) - (L=4)$  & 0.60 & -1.79 & -2.11 & -0.99\\[0.5ex]
\hline
\end{tabular}
\label{tab:table6}
\end{table}

From \cref{tab:table6} we observe that the largest improvement in WAIC is obtained by the mixture model for standardized errors with $L=2$. We compare the performance of our nonstationary GP emulator to the stationary GP emulator, TGP and CGP.

\begin{figure}[ht]
\begin{center}
\includegraphics[width=1\textwidth]{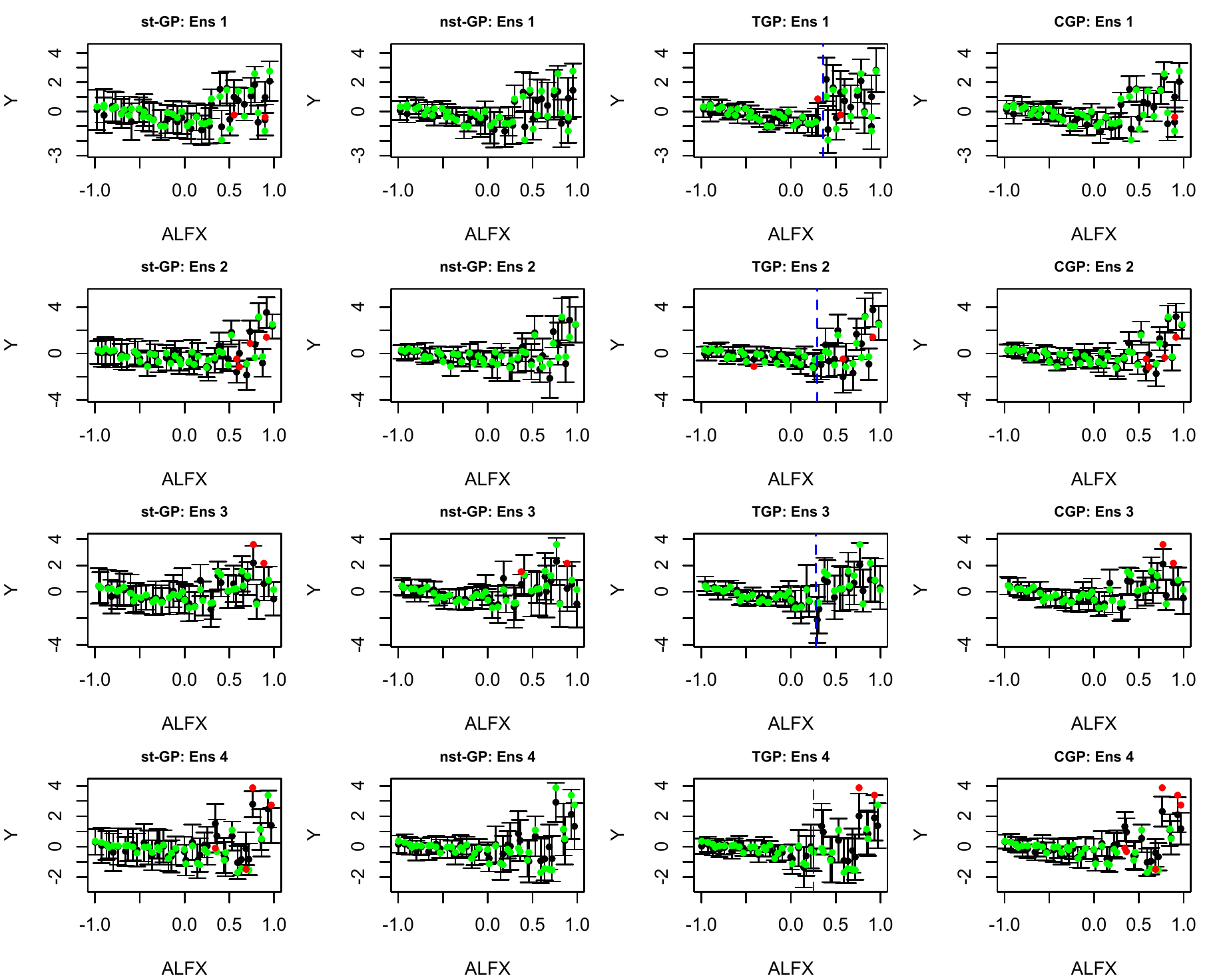}
\end{center}
 \caption{Comparison between stationary GP (\textit{st-GP}), nonstationary GP (\textit{nst-GP}), TGP, and CGP on modelling average potential temperature on four validation designs. Blue dashed lines correspond to partitions produced by TGP.
Each row is constructed by leaving one LHC out. The posterior mean and two standard deviation prediction intervals produced by emulators are in black. 
The green and red points are the model values, coloured `green' if they lie within two standard deviation prediction intervals and `red' if they lie outside. }
\label{fig:figure11}
\end{figure}

From \cref{fig:figure11} we observe that the stationary GP emulator fails to recognise the variability of the model response in relation to ALFX, i.e. the length of two standard deviation prediction intervals is the same across the whole range of ALFX. TGP demonstrates satisfactory performance for all four validation ensembles, however due to the hard partitioning into two separate regions mentioned in  \cref{sec:intro} the two standard deviation prediction intervals increase significantly for the ALFX$>$0. CGP performs well in the input region where the model is well-behaved, however is over-confident in the region where the model response varies the most, especially for sub-designs 2 and 4. Our nonstationary GP emulator demonstrates a gradual increase in the length of prediction intervals with increasing ALFX as we observe in the data. From \cref{table:7} we observe that our nonstationary GP model obtains the lowest Interval Score for ensembles 2 and 4 indicating good coverage, while \cref{table:8} demonstrates that our nonstationary GP emulator obtains the lowest RMSE for ensembles 1, 3 and 4.

\begin{table}[h!] 
\caption{Interval Score for the CNRM-CM5 experiments.}
\centering
\begin{tabular}{|c|c|c|c|c|}
\hline
Models & Ens 1 & Ens 2 & Ens 3 & Ens 4\\[0.5ex]
\hline 
st-GP &2.52 & 3.23 & 2.79 & 3.13\\
\hline
nst-GP & 1.99 & 2.09 & 2.57 & 1.97\\
\hline
TGP & 2.03 & 2.91 & 2.11 & 2.30\\
\hline
CGP &  1.91 & 2.38 & 2.32 & 3.73\\[0.5ex]
\hline
\end{tabular}
\label{table:7}
\end{table}

\begin{table}[h!] 
\caption{Root Mean Squared Errors (RMSE) for the CNRM-CM5 experiments.}
\centering
\begin{tabular}{|c|c|c|c|c|}
\hline
Models & Ens 1 & Ens 2 & Ens 3 & Ens 4\\[0.5ex]
\hline 
st-GP & 1.30 & 1.70 & 1.30 & 0.56\\
\hline
nst-GP & 1.28 & 1.66 & 1.26 & 0.54\\
\hline
TGP & 1.36 & 1.76 & 1.30 & 0.65\\
\hline
CGP &  1.38 & 1.65 & 1.26 & 0.60\\[0.5ex]
\hline
\end{tabular}
\label{table:8}
\end{table}

\section{Discussion}
\label{sec:disc}
In this paper we have introduced a nonstationary, diagnostic-driven, GP emulation of computer models via covariance kernel mixtures. We construct our nonstationary GP emulator in two steps. Firstly, we check if the stationary GP emulator performs well by considering plots of individual standardized errors against inputs in order to identify any signs of nonstationarity/heteroskedasticity \cite{Bastos2009}. We then fit a mixture model to the individual standardized errors to produce a mixture function prescribing the covariance kernel mixture for a nonstationary GP. We specify region-specific stationary covariance kernels, then establish the covariance kernel for our nonstationary GP as a mixture of these. The numerical examples together with the real-data application demonstrated the competitive performance of our method compared to the main nonstationary methods implemented in software, TGP and CGP.

A motivation for our approach is to mimic the approach we take to building emulators for computer models in practice. We may fit many such emulators to different model output over a session with our collaborators and use diagnostics from stationary fits to decide how to proceed. Our method treats these diagnostics as data that we then use to fit nonstationary models, making the fitting of many emulators more automatic, so that ultimately it can be done in house by the modellers themselves using our software.

There are a number of possible extensions to our developed methodology.  Firstly, we may remove the 2 stage approach altogether by operating with joint prior distribution $\pi(\bb{\beta}, \bb{\sigma}_L^2, \bb{\tau}_L^2, \bb{\delta}_L, \bb{\lambda}(\stanx)_L, L)$ using reversible jump MCMC \cite{Green1995, Kim2005} to attempt full Bayesian inference. However reversible jump MCMC could be time-consuming and operationally expensive, requiring a longer warm-up period \cite{Green1995, Pope2018}. Also this model specification could raise identifiability issues for $\stanx$, in particular to which input region $\stanx$ should be allocated. Intelligent choice of prior distribution in order to avoid confounding will be important here. Finally, we would be interested to further develop diagnostics for types of nonstationarity and tailor methods/priors to these.

\section*{Acknowledgments}
The authors gratefully acknowledge the support from Agence Nationale de la Recherche (ANR) (grant HIGH-TUNE \#ANR-16-CE01-0010). The authors would like to thank Romain Roehrig for providing us the ARPEGE-Climat SCM simulations output used in the present study. The authors also would like to thank the High-Tune project team for fruitful discussions.

\bibliographystyle{siamplain}
\bibliography{PaperRef}

\end{document}